# Plasmon-driven creation of magnetic topological structures

W. Al Saidi,[1] R. Sbiaa,[1*] Y. Dusch,[2] and N. Tiercelin[2]

[1]Department of Physics, College of Science, Sultan Qaboos University, P.O. Box 36, PC 123, Muscat, Oman

[2]University of Lille, CNRS, Centrale Lille, Université Polytechnique Hauts-de-France, UMR 8520 - IEMN, 59000 Lille, France

In the present research, we demonstrate the usage of plasmonic effects in thin film structures to control magnetic topological textures, specifically skyrmions and skyrmioniums. We investigate numerically the generation and alteration of these topological structures caused by hemisphere gold nanoparticle placed over a magnetic layer coated with a dielectric material. The electromagnetic and photothermal models are used to clarify the processes of producing heat and absorption, and the results were implemented in micromagnetic formalism to reveal the dynamics of magnetization under various conditions. Our findings demonstrate the significance of the laser pulse duration and the contact area between nanoparticles and the underlying magnetic layer in forming topological textures. In particular, we show how to generate a single skyrmion, multiple skyrmions, and skyrmioniums, and how to dynamically transition between these states. These results highlight the possibility of manipulating magnetic textures by using plasmonic effects, which presents significant opportunities for spintronics and non-conventional computer applications.

# I. INTRODUCTION

Skyrmions are stable magnetic structures where the spins rotate around their core in a swirling configuration [1]. These nanoscale magnetic textures have interesting properties making them promising candidates for next-generation technologies [2–8]. They exhibit high topological stability, can be displaced with a low current density and in contrast to magnetic domain walls are less sensitive to material defects. Skyrmions are a type of magnetic topological textures that are characterized by their topological charge $Q$ [9]. Like in the case of particles and antiparticles, the existence of a skyrmion with $Q = +1$ can be accompanied by an antiskyrmion with an opposite chirality. A skyrmionium, which is another topological object is formed by an antiskyrmion in the core of a skyrmion leading to $Q = 0$. Although skyrmions have been intensively studied, skyrmioniums have been recently the focus of several reports motivated by the non-existence of Magnus force as a consequence of the vanishing topological charge $Q$ [10–14].

Before they can be moved, creating topological textures is the first key step for their implementation in functional spintronic devices. Several methods were proposed for creating skyrmions, but only a few have been designated for the case of both skyrmions and skyrmioniums [11,13,15,16].

In this study, we numerically explored the plasmonic effect in creating either skyrmions or skyrmioniums by finite elements methods. We considered gold nanoparticles (NP)s with different shapes deposited on $SiO_2$ thin film. The magnetic layer is a thin Co film capped against oxidation with a Pt layer, onto which is deposited a dielectric layer. By employing pulsed laser irradiation at resonance wavelengths, we establish a precise and controlled method for inducing localized heat, facilitating the formation of different topological textures. Depending on the contact area between the NP and the insulator, it was possible to create either a single skyrmion, more than one skyrmion or a skyrmionium in the range of ps. Furthermore, the plasmonic effect can be used to transform a skyrmionium into a skyrmion using a large laser pulse of about 1.5 ns. Merging plasmonics and skyrmionics opens avenues for advanced applications in race track memory and non-conventional computing.

## II. THEORETICAL MODELS

For an optimal plasmonic effect, we examined the effect of the contact area between a gold nanoparticle (NP) and an insulator deposited on the top of a magnetic layer. The contact area was determined by considering the facing diameter f which in this study was varied from 25 nm to 55 nm. Figure 1(a) is a schematic representation of a working device where the skyrmion can be created at one pad by plasmonic effect: Compared to a more classical skyrmionic device, several gold nanoparticles are added on top of the nucleation pad. It can be noted that in a practical device, the nanoparticles can be deposited by self-assembly, chemical or physical methods. The laser-assisted creation process of magnetic skyrmions or skyrmioniums will be discussed later. By spin transfer torque, these topological textures will be displaced inside the devices and others will be created if needed. Although the laser spot is much larger than the gold NPs size (diffraction limit), skyrmions with size even smaller than the NP can be created. Furthermore, the size and the number of skyrmions can be well controlled by the distribution and size of NPs. To ensure a consistency in the calculation, the volume $V$ of the NP was kept constant as evaluated from a half sphere with 40 nm diameter while the facing diameter f was varied. Figures 1(b) and (c) show two nanoparticles with f = 25 nm and 55 nm, respectively. In the first case, the NP is a portion of a sphere with the bottom part truncated, indicative of a missing portion. For f = 55 nm, shown in Fig. 1(c), the gold nanoparticle represents only the top part of a sphere, resulting in a wider appearance compared to the first case. In addition to these cases, we have also investigated the cases of f = 40 and 45 nm. By changing the shape of the nanoparticle, we aim to evaluate the impact of the contact area on the plasmonic effect which will be later included into the micromagnetic model. It has been demonstrated that the shape and size of NPs can be controlled by changing the temperature during the growth or the synthesis [17–19] which is a crucial factor to take into account when designing the properties of nanoparticles for a specific applications.

## II. 1. ELECTROMAGNETIC MODEL

A plane wave illuminated the nanoparticle along the *z*-axis, and the azimuthal angle $\varphi$ and angle of incidence $\theta$ controlled the electric field polarization. In order to decrease artifacts and prevent reflections, perfectly matched layer domains and port boundary conditions were used. Periodic boundary conditions were used in the calculation to ensure stability and prevent edge effects by modeling an infinite thin film. The amount of absorbed electromagnetic radiation is re-emitted as heat. Therefore, in thermoplasmonics, absorption is the most important factor [20] which can be calculated using the Poynting theorem [21]

$$\sigma_{abs} = \frac{n_m k \varepsilon''_{NP}}{E_o^2 \varepsilon_m} \iiint_{V_{NP}} |E(r)|^2 dV \qquad (1)$$

In Eq. (1), the absorption cross-section $\sigma_{abs}$ was computed while taking the incident wave energy, material characteristics, and refractive index into account. The gold NP is on top of a thin film that consisted of the following layers: $SiO_2$(1 nm), Pt(1 nm), Co(10 nm), Ta(10 nm) [22–24], and $SiO_2$ substrate of 100 nm, which were stacked in order from the top. Pt is used as a protective layer against oxidation and Ta is a buffer layer for both a good adhesion of magnetic layer and for better growth. It is worthy to note that we used a magnetic layer that could be Co or any other magnetic multiayers that favour the formation of skyrmions or magnetic topological texture [25–29].

We identified the resonance wavelength of the gold NP, which indicates maximal absorption and effective heat dissipation upon interaction with incident light, by examining the absorption cross-section for different wavelengths.

## II. 2. PHOTOTHERMAL MODEL

The heat-transfer analysis was carried out numerically in the current study and is expressed as follows:

$$\partial T/\partial t + U \cdot \nabla T = k/(\rho C_p) \nabla^2 T + q \qquad (2)$$

where $T$, $t$ and $\vec{U}$ stand for the temperature, time, and the velocity vector, respectively. The parameters $k$, $\rho$ and $C_p$ are the fluid thermal conductivity, density, and heat capacity, respectively. The parameter $q$ is the local heat generated per unit time and is derived from $\sigma_{abs}$ expressed in Eq. (1) under a specific laser irradiance intensity and wavelength. The local heat generation term can be written as

$$q(r) = \frac{1}{2}\omega \Im[\varepsilon_{NP}(\omega)]|E(r)|^2 \qquad (3)$$

where $\varepsilon_{NP}$ is the nanoparticle dielectric constant, $\omega$ is the laser frequency and $E$ is the electric field.

## II. 3. MICROMAGNETIC FORMALISM

To study the magnetic behavior of a thin film in the presence of the plasmonic effect induced by gold nanoparticles, micromagnetic simulations were conducted [30]. The calculation model was designed to comprehensively explore the magnetization dynamics and magnetic properties of the system under various conditions. The grid size of 512 × 512 × 1 and cell dimensions of 1 nm × 1 nm × 1 nm were used. To minimize boundary effects and enhance the accuracy and real-world relevance of the simulations, periodic boundary conditions were implemented in both the $x$ and $y$ directions. The intrinsic properties of the magnetic material such as the saturation magnetization, the exchange stiffness, the interfacial Dzyaloshinskii-Moriya strength, the uniaxial magnetic anisotropy and the damping constant were fixed to 600 kA/m, $7\times10^{-12}$ J/m, 2.0 mJ/m$^2$, $0.67\times10^6$ J/m$^3$ and 0.1, respectively. External factors, such as predefined magnetic fields, current density, and distributed heat extracted from electromagnetic and photothermal models were included in the modified Landau Lifshitz-Gilbert equation.

## III. RESULTS AND DISCUSSION

The efficiency of plasmonic effect can firstly be seen on the absorption $\sigma_{abs}$ than on the temperature profile which will be discussed later. Fig. 1(d) shows the absorption spectra for the four different facing diameters f. The highest absorption of about $8.6 \times 10^{11}$ nm$^2$ at a wavelength of 1410 nm is observed for f = 55 nm configuration; *i.e.* larger contact area between the nanoparticle and the underneath insulator. The resonance wavelength $\lambda_R$ shifts to 1215 nm as the face diameter drops to 45 nm, supporting the Mie theory trend [31]. Remarkably, the resonance appears at 1045 nm for f = 40 nm and shifts to 745 nm for f = 24 nm where the absorption decreased to $4.72 \times 10^{11}$ nm$^2$. According to Mie's theory, the absorption cross-section of an NP is directly related to its size and shape, with larger particles generally exhibiting increased absorption cross-sections, there is a red-shift in resonance towards longer wavelengths, which implies that larger particles have better absorption efficiency. Additionally, the plasmonic resonance characteristics, including the absorption cross-section, depend on the size, shape, composition, and the local environment of the nanoparticles. Therefore, in the case of a complete sphere and a hemisphere with constant volume, the sphere is expected to exhibit a larger absorption cross-section compared to the hemisphere due to its larger size and surface area. This pattern highlights the plasmonic effects is morphology dependent as can be seen in Fig. 1(d). The collective oscillation of free electrons at the nanoparticle surface significantly increases the local electric field when it is illuminated by the incident light. This behavior is most noticeable when the electric field is amplified by orders of magnitude in comparison to the incident light in the surrounding area of the NP. In the context of our investigation, we have explored the intricate behavior of electric field distribution as a function of position $x$ directly below gold NP at the interface with the thin film, presented in Fig. 2(a). The $SiO_2$, Pt, Co and Ta make up the thin film structure investigated including the $SiO_2$ substrate. It is clear that the electric field is concentrated at the lateral sides of the NP borders for all cases at the resonance wavelength. The curvature and symmetry play a crucial role in influencing the distribution of surface plasmons, leading to localized electric field enhancements at distinct regions. In the case of hemisphere and spherical caps, the lateral sides exhibit the highest electric field intensity due to the curvature and symmetry of the structures. Furthermore, the presence of a dielectric substrate environment leads to field enhancement due to localized surface plasmon resonances (LSPR) and reflections. These reflections contribute to the constructive interference of the light waves with a further enhancement of the electric field at the lateral sides of the nanoparticle. As a result, the combination of curvature, symmetry, LSPR and reflections at the interface leads to the maximum intensity of the electric field at the lateral sides of the hemisphere. The distribution of electric field for f = 25, 40, 45 and 55 nm is presented in Figs. 2(c)-(f), respectively. From Fig. 2(a), the electric field values are reduced when the facing diameter is 24 nm compared to cases with larger values, indicating that the electric field intensity can be controlled by the size of the contact area between the NP and the capping insulator layer. The enhancement of the electric field at plasmonic resonance enhances the absorption and scattering cross-sections of the nanoparticle, leading to the generation of current density. The regions where the current density is maximized are observed at lateral sides due to curvature at the edges as can be seen in Fig. 2(b). These areas create "hotspots" where the electric field is strong due to plasmonic effect. The presence of a dielectric substrate can further enhance the current density produced at plasmonic resonance [32]. Moreover, the interaction between the NP and the surrounding environment can lead to the generation of spin currents by surface plasmon resonance that contributes to the increase in current density at plasmonic resonance [33]. Figures 2(g-j) present the distribution of current density $J$ along the $z$-axis in the $x$-$y$ plane for different facing diameters of gold NP. The current density provides an insight into the flow of electric current through the nanoparticle structures, offering a visual representation of how different facing diameters influence the spatial distribution of current. It can be clearly seen that $J$ increases by a factor of 3 as the contact area becomes larger. Similarly to the electrical field, the current density is stronger at the edges of the nanoparticles. The increased electric field causes current density near the NP, which in turn causes electron flows. As a result of the small size of gold NP and high current density, the heat produced by the Joule effect is localized around it. The temperature of the NP and its surroundings rises as a result of the heat produced by the Joule effect. While the non-uniform electric field induces spatially varying heat generation rates within the NP, its higher thermal conductivity than the surrounding medium facilitates instantaneous spread of the heat. The total heat generation can be expressed as the product of the absorption cross-section $\sigma_{abs}$ and the intensity of the incident light. In our case, a picosecond pulsed laser illuminated with an intensity of 142 mW/mm [34] for a duration of 500 ps. To reduce the risk of thermal damage to any of the layers in the investigated stack, especially the magnetic layer, a relatively low laser intensity, similar to that of magnetic recording media, is used. For different configurations, the calculated temperature in the magnetic layer as a function of time is plotted in Fig. 3(a). The observed differences of the temperature characteristic are caused by variances in the distribution of the electric field and $\sigma_{abs}$ around the gold nanoparticles. Because of plasmonic effects, larger facing diameters, f = 55 nm had a maximum temperature of ~ 680 °C, where it showed a greater absorption cross-section and enhanced electric field strength. This leads to more efficient light absorption and greater heat generation in the magnetic layer. On the other hand, the maximum temperature obtained for smaller contact area (f = 24 nm) is only about 320 °C for smaller facing diameters (f = 24 nm) due to relatively less absorption and smaller electric field enhancement.

The temperature distribution as a function of the position *x* within the magnetic layer presented in Fig. 3(b), revealing a Lorentzian characteristic. The temperature profile along the *x*-axis shows a distinct peak temperature at the center that is gradually attenuated towards the edges. As expected from the discussion above, the maximum temperature $T_M$ shows an almost linear behaviour with the parameter f, corresponding to the contact area. The spatial temperature distribution in both *x* and *y* axis is presented in Fig. 3(c)-(f) for a facing diameter of 25 nm, 40 nm, 45 nm and 55 nm, respectively. For larger value of f, a broader temperature profile, reflecting a more extended influence of heat within the magnetic layer can be observed.

After the calculation of the temperature profile originating from plasmonic effect, we conducted micromagnetic calculation to understand the magnetization behaviour. In the LLG equation, the temperature dependence on time was included for each cell of the sample as discussed in Section II.

The plasmonic effect from gold NPs influences the magnetization dynamics in the thin film. When a pulsed laser is applied to the whole stack at the resonance wavelength, a localized heat in the vicinity of the NPs is generated by plasmonic effect. This heat, in turn, causes the magnetic moments to fluctuate, leading to the formation of skyrmions for the cases of f = 40, 45 and 50 nm. However, the absence of skyrmions in the case of f = 24nm configuration is noteworthy. At a facing diameter of 24 nm, the NP configuration might not induce sufficient plasmonic effects or localized heat to initiate the skyrmion formation. The lower curvature and reduced surface area of the NP may result in less efficient plasmonic energy conversion. The size and shape of the NP may influence the heat distribution and, consequently, the magnetic response.

In addition to the formation of a skyrmion at the NP position, we noticed a difference in the creation time. For NPs with a facing diameter of 50 nm and 45 nm, the skyrmion could be created at about 35 ps and 65 ps, respectively indicating that larger contact area leads to a faster skyrmion creation. At this stage of the calculations, we did not observe a change of skyrmion size for these cases with a skyrmion diameter of about 30 nm. In order to investigate this phenomenon, we looked into how the duration of the pulse affected the magnetic texture induced by gold NPs. We calculated the magnetic spins evolution on time under different pulse durations $\tau$. Figure 4 shows a distinctive case where the manipulation of $\tau$ exhibits a profound influence on both the topological characteristics and the quantity of generated skyrmions. Figure 4(a) illustrates the computed *z*-component of magnetization $m_z$ as a function of time for three different pulse widths; 50 ps, 100 ps, and 150 ps. For a laser pulse width of 50 ps, two well separated skyrmions were created by plasmonic effect, characterized by a topological charge $Q = 2$, as depicted in Fig. 4(b). The evolution of $Q$ with time shows rapid fluctuations on top of a steady increase with temperature. Topological stabilization events can be identified by a sudden balance of the net local topological charge. The subsequent Figs. 4(c-f) are snapshots showing the magnetic moment distribution at $t$ of 25 ps, 60 ps, 90 ps and 1 ns, respectively. From approximately 60 ps, one can see the beginning of the separation of two magnetic regions, which ends up by the formation of two stable skyrmions with a small size of about 30 nm. The separation of these two skyrmions is due to the repulsive interaction between them.

For a pulse width of 100 ps, a skyrmionium with a null topological charge is formed as shown in Fig. 4(b) and described in the snapshots of Figs. 4(g-j). The outer skyrmion is deformed into a ring by the inner skyrmion, which has an inverted polarity. This configuration is characterized by a swirling spin texture with no net winding, resulting in a distinct donut-shaped magnetic structure. The nucleation of skyrmioniums is presented in Fig. 4(g) where a small antiskyrmion is created within a relatively large domain. As time elapses, a clear formation of skyrmion can be seen in the selected snapshots of Fig. 4(g-j) taken at $t$ equal to 130 ps, 200 ps, 260 ps and 1 ns, respectively. Furthermore, a pulse width of 150 ps resulted in the creation of a single skyrmion characterized by a topological charge of +1 [Fig. 4(b)]. A skyrmion with a topological charge of +1 represents a standard skyrmion configuration. The size of the two combined skyrmions [Fig. 4(f)] and a single skyrmion [Fig. 4(n)] have a diameter of about 30 nm, notably smaller than the contact diameter of the gold NP. The skyrmions size remained unaffected by variations in pulse width. Contrary to expectations, changes in the duration of laser pulses did not lead to discernible alterations in the size of the skyrmionic structures. Instead, our findings suggest that the primary determinants of skyrmion size reside in the inherent material parameters and characteristics [35]. Figures 4(k-n) are images of the magnetic moments around the nanoparticle for a pulse width of 150 ps and recorded at 50 ps, 300 ps, 600 ps and 1 ns, respectively. Through careful modulation of the laser pulse width, particularly in the case of a gold NP with a 50 nm diameter, precise control over the topological properties and the number of resulting skyrmions is achieved. It can be seen from Fig. 4(f), (j) and (n) that the size of skyrmions are similar and smaller than the nanoparticle size while skyrmionium has a larger size.

Up to this point, heat changes have only affected the shape, not the existence, of skyrmion. As discussed before, it is possible to generate a single skyrmion, two separated skyrmions and a skyrmionium by adjusting the laser pulse width as shown in Fig. 4. In the last set of simulations, we tried to understand whether it is possible to convert a skyrmionium to a skyrmion using the plasmonic effect. For this purpose, we investigated the impact of a second laser pulse on an existing skyrmionium created earlier. Through irradiation with a 70 ps long laser pulse and based on plasmonic effect as explained earlier, a conversion of the initial skyrmionium to a skyrmion was possible as can be seen in Fig. 5. Different pulse durations result in different energies and heating profiles within the magnetic material as can be seen in Fig. 3(a) when time is of few tens of ps. The observed transformation dynamics highlight the sensitivity of the magnetic topological textures to external stimuli, showcasing the ability of specific pulse durations to modify the nature of a pre-existing magnetic topology. Additionally, at 150 ps we observed that varying pulse durations could drive the conversion of a skyrmionium to two distinct skyrmions (not shown). For instance, pulse widths of certain durations, such as 140 ps, demonstrated the capacity to transform a skyrmion into a skyrmionium. This dynamic interplay between pulse characteristics and magnetic configurations underscores the rich and tunable nature of *plasmonically*-induced transformations of magnetic textures. In parallel to our findings, the study by Khela *et al.* [36], show that ultrafast laser causes phase transitions between various topological spin textures in two-dimensional van der Walls magnets. Furthermore, recent study achieved conversion of a skyrmion to skyrmionium and vice versa through controlled manipulation using square pulses along the *x*- direction while maintaining an applied magnetic field perpendicular to the sample surface [37]. This convergence of findings suggests that both plasmonic, laser-driven and pulsed current with constant magnetic field stimuli hold promise for reconfigurable topological textures.

## IV. CONCLUSION

In conclusion, this work analyzes the creation and manipulation of magnetic topological structures in thin film using plasmonic effects. Based on the strong electromagnetic fields and heat generated around the nanoparticles, it is possible to create different magnetic textures. Depending on the laser pulse duration and the contact area between the magnetic thin film and the nanoparticle, one can create a single skyrmion, multiple skyrmion and even skyrmionium. This method when used with an array of nanoparticles is very effective to create a large number of magnetic textures with low energy due to plasmonic effect. Furthermore, the formation of these stable magnetic topological objects occurs in the range of picoseconds.

**Acknowledgements**

RS would like to acknowledge the support from the HMTF Strategic Research of Oman (grant no. SR/SCI/PHYS/20/01).

**Author contributions**

R.S. conceived the idea and analyzed the data. W.S. conducted the simulation and interpreted the results. YD and NT contributed to the interpretation of the results. All authors discussed the results and provided inputs to the manuscript.

**Competing interests**

The authors declare no competing interests

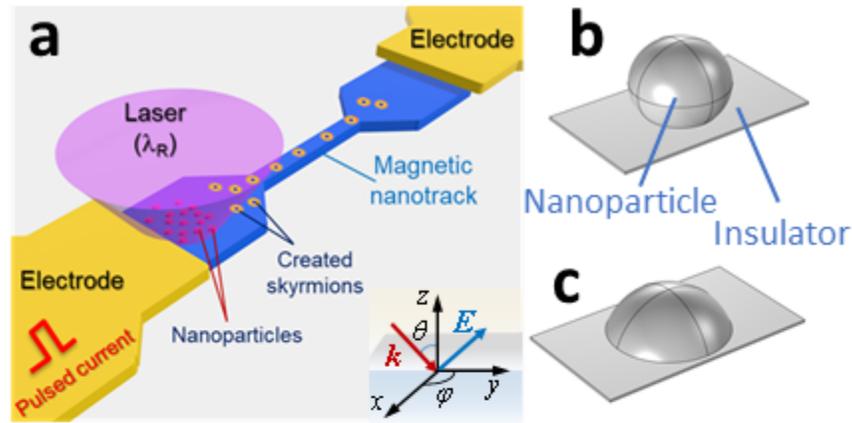

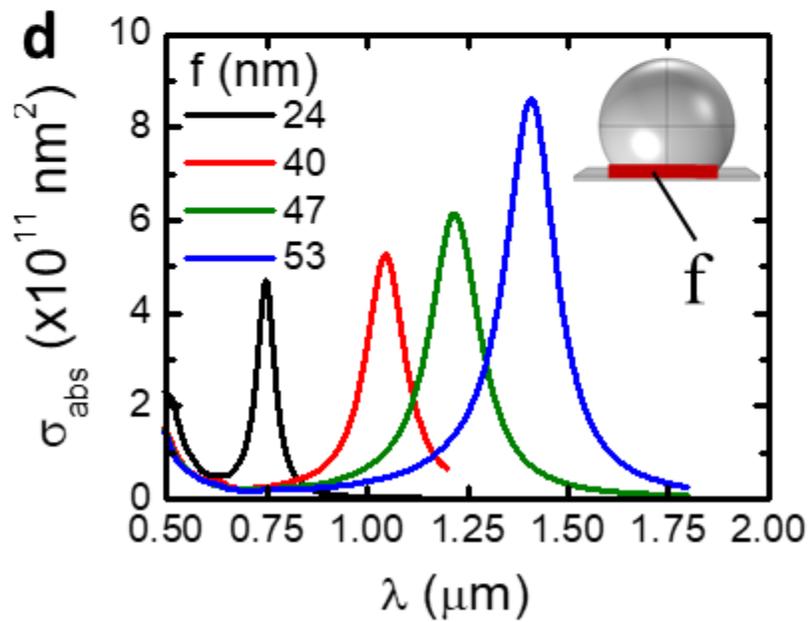

FIG.1. (a) Schematic representation of the proposed trajectory device design for skyrmion creation using plasmonic effects and manipulation using pulsed current. (b) and (c) are the gold NP spherical cap shapes with facing diameters (f) of 25 nm and 55 nm, respectively. (d) The calculated absorption cross-section spectra for different facing diameters (f = 25, 40, 45, and 55 nm) with a constant volume as a function of wavelength.

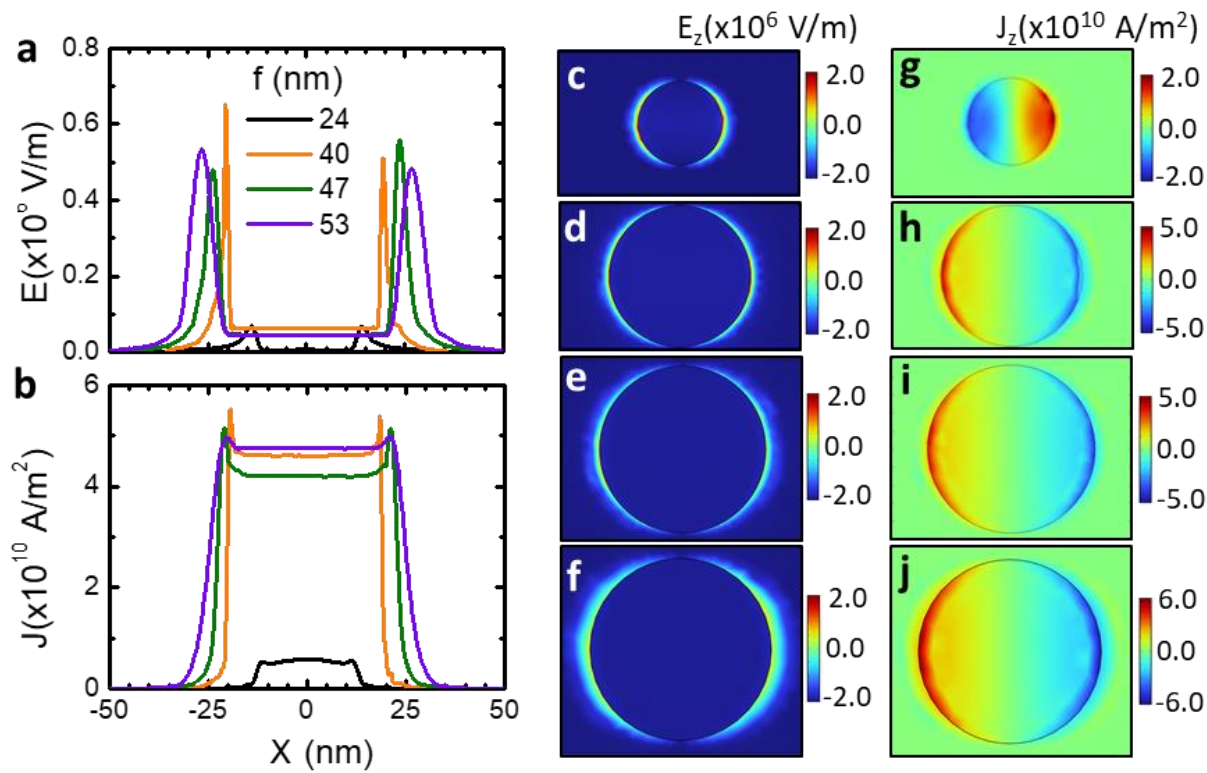

FIG. 2. (a) Electric field distribution along the x-axis for different facing diameters (f) of AuNP (f = 25, 40, 45, and 55 nm). (b) Current density in the z-axis as a function of x for various facing diameters (f = 25, 40, 45, and 55 nm). (c) Electric field distribution in the *x-y* plane for a gold nanoparticle with a facing diameter of 25 nm (f = 25) (d) f = 40 (e) f = 45 and (f) f = 55. (g) Current density distribution in the *z*-axis in the *x-y* plane facing diameter of 24 nm (f = 25). (h) f = 40 (i) f = 45 and (j) f = 55 nm.

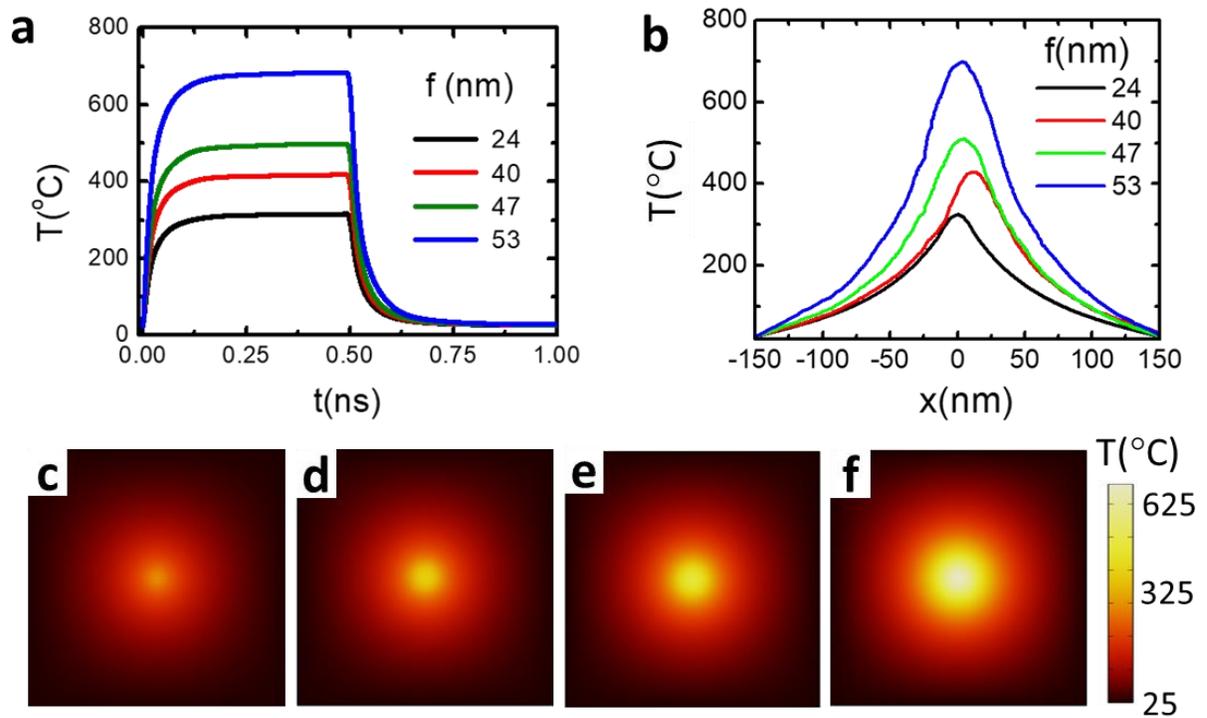

FIG. 3. (a) Temperature (T) versus time for different facing diameters (f = 25, 40, 45, 55 nm) under pulsed laser irradiation with a duration of 500 ps and intensity of 142 mW/mm. (b) The temperature versus the position $x$ within the magnetic layer for various facing diameters. (c)-(f) in-plane representation with a color-coded temperature profile for f = 25 nm, 40 nm, 45 nm and 55 nm, respectively. The temperature profile is taken at $t$ = 500 ps.

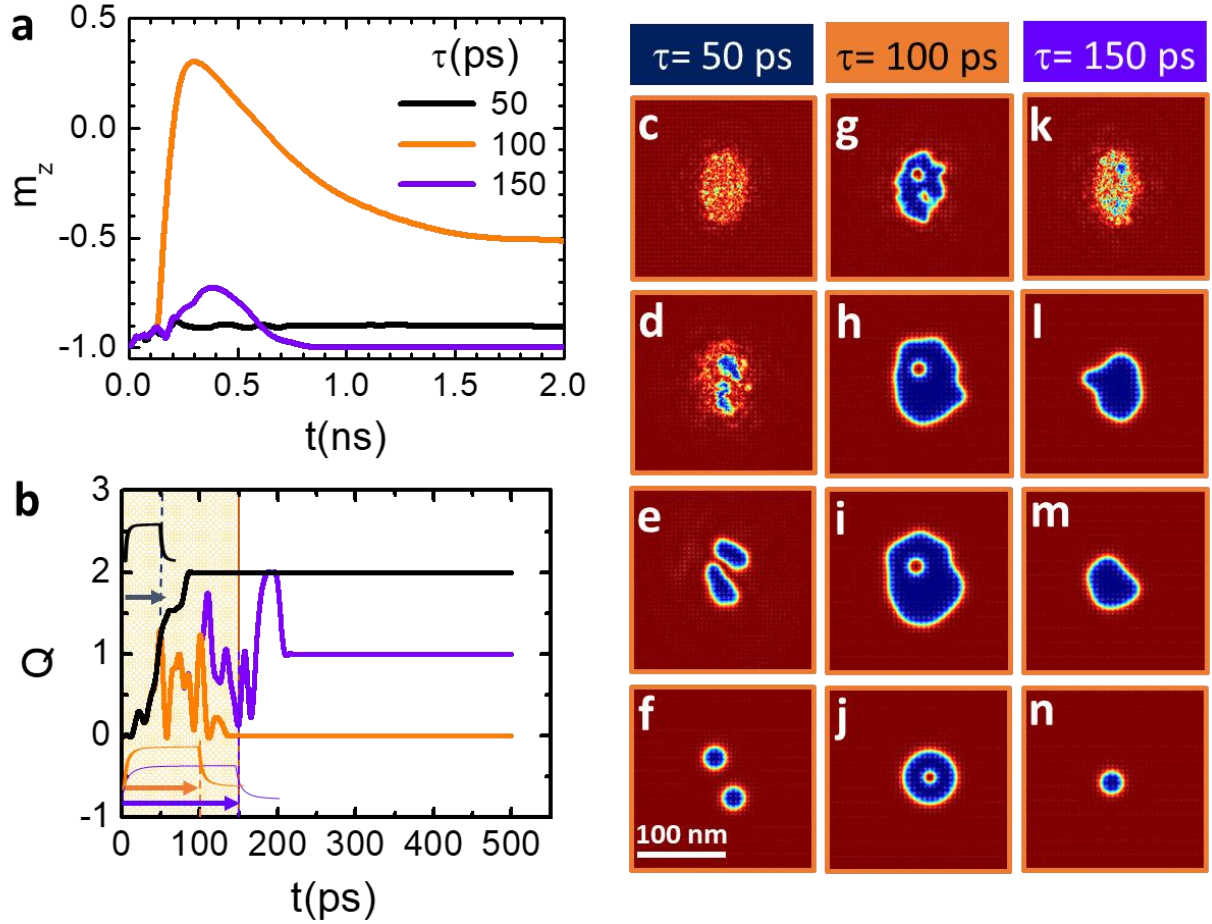

FIG. 4. (a) The out-of-plane component of the magnetization $m_z$ versus time for pulse durations of 50, 100, and 150 ps for the case of f = 55 nm. (b) Topological charge ($Q$) versus time for different pulse durations. (c)-(f) Nucleation process for pulse width 50 ps, showing the formation of two skyrmions at $t$ of 25 ps, 60 ps, 90 ps and 1 ns, respectively. (g)-(j) Nucleation process for pulse width 100 ps, showing the formation of a skyrmionium at $t$ of 130 ps, 200 ps, 260 ps and t = 1 ns, respectively. (k)-(n) Nucleation process for pulse width 150 ps, showing the formation of a skyrmion at $t$ = 50 ps, 300 ps, 600 ps and 1 ns, respectively.

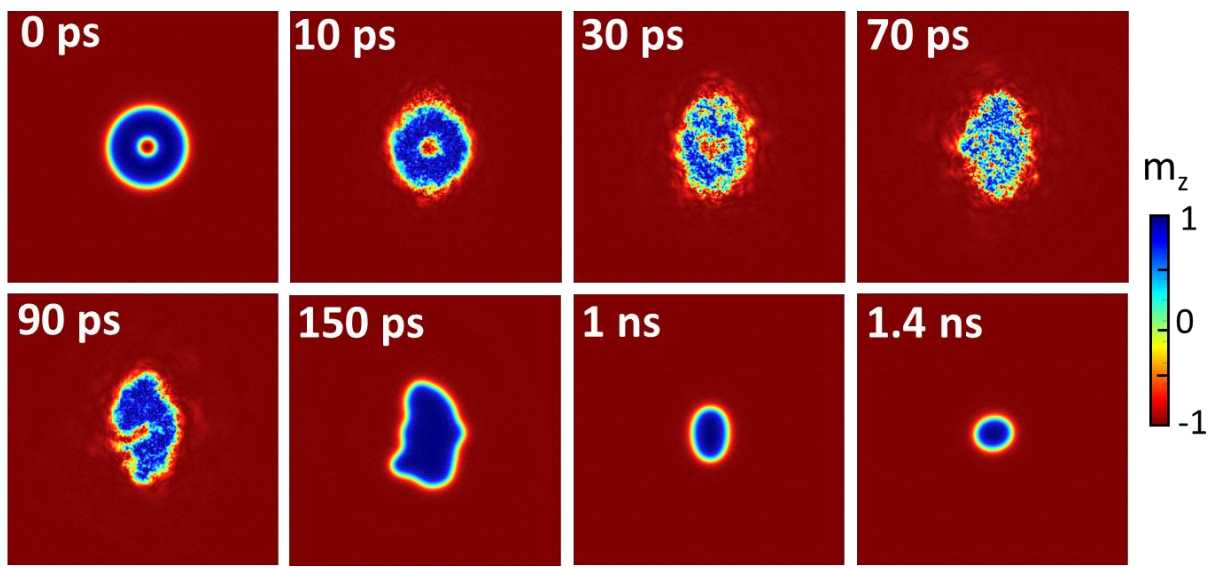

FIG. 5. Time-Resolved transformation of skyrmionium to skyrmion under plasmonic effect using a pulse duration of 70 ps.